\documentclass[prd,aps,twocolumn,floats,floatfix,nofootinbib]{revtex4-1}

\usepackage[utf8]{inputenc}
\usepackage{amssymb,bm}
\usepackage{graphicx}
\usepackage{dcolumn}
\usepackage{bm}
\usepackage{graphics}
\usepackage{slashed}
\usepackage{enumerate}
\usepackage{amsmath}
\usepackage{url}
\usepackage[usenames,dvipsnames,svgnames,table]{xcolor}
\usepackage{color}
\usepackage{xcolor}

\definecolor{color1}{RGB}{191, 0, 255}

\begin{document}

\title{Universality of the turbulent magnetic field in  hypermassive neutron stars produced by binary  mergers} 
  
\author{
Ricard Aguilera-Miret$^{1,2,3}$,
Daniele Vigan\`o$^{4,3,2}$,
Carlos Palenzuela$^{1,2,3}$
}

\affiliation{${^1}$Departament  de  F\'{\i}sica,  Universitat  de  les  Illes  Balears,  Baleares  E-07122,  Spain\\
$^2$Institute of Applied Community and Community Code (IAC3),  Universitat  de  les  Illes  Balears,  Palma  de  Mallorca,  Baleares  E-07122,  Spain\\
$^3$Institut d'Estudis Espacials de Catalunya (IEEC), 08034 Barcelona, Spain\\
$^4$Institute of Space Sciences (IEEC-CSIC), 08193 Barcelona, Spain}
 
\begin{abstract}
The detection of a binary neutron star merger in 2017 through both gravitational waves and electromagnetic emission opened a new era of multimessenger astronomy. The understanding of the magnetic field amplification triggered by the Kelvin-Helmholtz instability during the merger is still a numerically unresolved problem because of the relevant small scales involved. One of the uncertainties comes from the simplifications usually assumed in the initial magnetic topology of merging neutron stars. We perform high-resolution, convergent large-eddy simulations of binary neutron star mergers, following the newly formed remnant for up to $30$ milliseconds. Here we specifically focus on the comparison between simulations with different initial magnetic configurations, going beyond the widespread-used aligned dipole confined within each star. The results obtained show that the initial topology is quickly forgotten, in a timescale of few miliseconds after the merger. Moreover, at the end of the simulations, the average intensity ($B\sim 10^{16}$ G) and the spectral distribution of magnetic energy over spatial scales barely depend on the initial configuration. This is expected due to the small-scale efficient dynamo involved, and thus it holds as long as: (i) the initial large-scale magnetic field is not unrealistically high (as often imposed in mergers studies); (ii) the turbulent instability is numerically (at least partially) resolved, so that the amplified magnetic energy is distributed across a wide range of scales and becomes orders-of-magnitude larger than the initial one.
\end{abstract}

\maketitle

\section{Introduction}
The GW170817 event~\cite{abbott17a,abbott17b} was arguably one of the most important astrophysical findings of the last decade for several reasons. First, it demonstrated that binary neutron star (BNS) mergers can produce strong gravitational wave (GW) signals and power bright electromagnetic emissions on a broad range of the spectrum~\cite{goldstein2017,savchenko2017,abbott17d,abbott17c,metzger17,davanzo2018,fong2019,dobie2018,mooley2018}. Second, the comparison of theoretical models with the observations allowed to narrow some of the physical properties of neutron stars (NSs) such as their radius, maximum mass, tidal deformability and equation of state (EoS) (see, e.g.,~\cite{margalit17,shibata2017modeling,abbott2018}). In addition, it also served to measure the GW speed with unprecedented precision, setting strong constraints in many alternative theories of gravity (see, e.g.,~\cite{PhysRevLett.119.251301,2018EPJC...78..738G}). 

Although the dynamics of BNS mergers is fairly well understood (e.g., \cite{ciolfi2020key}), there are still many open questions regarding the details of the physical processes taking place during and after the merger. Here we focus in one of them, namely the amplification and large-scale reorganization of the magnetic field, which is thought to be necessary in order to launch a magnetically dominated jet associated to the short gamma-ray burst (SGRB) (see e.g.~\cite{Mckinney2009,10.1093/mnras/staa955}). Although the BNS merger and post-merger evolution of the remnant has been extensively studied through general-relativistic magnetohydrodynamics (GRMHD) simulations~\cite{palenzuela13a,kiuchi14,neilsen2014magnetized,kiuchi15,giacomazzo15,palenzuela15,ruiz16,kiuchi18,ciolfi2019,ciolfi2020collimated,ruiz2020,mosta2020}, the problem is not fully resolved yet. The impossibility of capturing the small (but dynamically important) scales induced by the MHD instabilities at play, possibly combined with other instabilities, obscures a definite answer on what is the topology and intensity of the magnetic field in the remnant, when the turbulent amplification approaches saturation. Moreover, these strong magnetic fields are thought to enhance the angular momentum transport (see e.g.~\cite{ciolfi2020key} and references therein), a key factor in the fate of the remnant. In presence of large-scale magnetic fields, the formation of the jet can be favored, and the amount mass ejecta could also change, compared to a non-magnetized remnant. Moreover, the presence of a jet or magnetically driven winds also depends on the field topology.

Different GRMHD simulations have attempted to resolve all the relevant scales of the problem. The highest-resolution simulations so far~\cite{kiuchi18} showed that the average magnetic field of the remnant can be amplified from $10^{13}$G to $10^{16}$G during the first miliseconds after the merger. Unfortunately, even the very fine spatial grid size used there (i.e. $\Delta \sim 12.5$m) is still much larger than the estimated wavelength of the fastest-growing unstable modes of the Kelvin-Helmholtz instability (KHI): insufficient to well capture the small, highly dynamical scales and the magnetic field amplification at this stage. As a result, the saturation level of the magnetic field intensity was not converging to any clear value.

Nowadays (and arguably in the foreseeable future) it is not possible to perform direct numerical simulations of this scenario, since the range of the relevant scales (from hundreds/thousands km of the domain to the sub-meter shear layer thickness) makes the computational cost unfeasible, even employing the most efficient numerical and parallelization methods currently available. In order to overcome this limitation, different strategies have been implemented to reproduce the under-resolved amplified small-scale magnetic field. A commonly used one is to impose a purely poloidal magnetic field, with unrealistically large strengths $\sim10^{14-16}\,$G, either before or after the merger(e.g.,~\cite{ruiz16,kiuchi18,ciolfi2019,ciolfi2020collimated,ruiz2020,mosta2020}). This choice is hardly comparable to the real effects of the dynamo operating at small scales, for which purely large-scale ordered field is not an expected outcome.

Other alternatives involve the use of large-eddy simulations (LES), which consist on including extra terms in the {\em discretized} version of the evolution equations to account for the unresolved sub-grid scale (SGS) dynamics (see e.g.~\cite{zhiyin15}). The main idea of the LES is to reproduce the imprints of the sub-grid dynamics on the large-scale (numerically resolved) fields, thus providing a magnetic field growth with a realistic topology and spectrum. Following this line, we have recently extended and implemented the so-called gradient SGS model used in non-relativistic fluid dynamics~\cite{leonard75,muller02a} to the non-relativistic~\cite{vigano19b}, special~\cite{carrasco19} and general relativistic~\cite{vigano20} MHD with excellent results on capturing the small scale effects of turbulent flow. We have performed LES of BNS coalescence~\cite{aguilera20}, and found that the average magnetic field in the remnant is amplified with much less computational resources than the higher-resolution simulations leading to comparable results. In an accompanying paper~\cite{palenzuela21}, we show that high-resolution LES provide an amplification of the average magnetic field from $10^{11}$G to $10^{16}$G. More importantly, for the first time the magnetic field strength and its energy spectral distribution are converging to the same saturated level.

The work shown in this letter, which uses the same techniques as in~\cite{palenzuela21}, focuses on the role of the initial topology and intensity of the magnetic field in the post-merger remnant. Most (if not all) simulations of magnetized BNS mergers up to date start with mainly dipolar magnetic fields, often confined in each star, for simplicity. However, magnetic topologies are expected to be much richer, with relevant contributions from small scales and from the magnetospheric currents. This holds throughout a neutron star's life: at birth after the core-collapse supernova~\cite{reboul-salze21}, for middle (Myr) ages (as shown by magnetars' observations~\cite{tiengo13,borghese15} and simulations~\cite{gourgouliatos16}) and for late (Gyr) ages similar to what merging NSs should have (as shown by NICER studies of old millisecond pulsars~\cite{riley19}). The key question we want to address here is the following: how the choice of the pre-merger configuration affect the final magnetic field of the remnant? The results presented in this letter indicate that the memory of the initial magnetic field configuration is lost during the amplification phase induced by the small-scale dynamo. For all the initial topologies considered, the bulk of the remnant (i.e., the regions with $\rho \geq 10^{13} ~\rm{g~cm^{-3}}$) is endowed with a very similar isotropic, turbulent-like configuration with an average magnetic field of approximately $10^{16}$G.

The paper is organized as follows: the evolution equations and the numerical setup are described briefly in \S\ref{sec:evol_and_setup}. The results of the simulations are presented and analysed in \S\ref{sec:results}. Conclusions are drawn in \S\ref{sec:conclusions}.

\section{Initial models}\label{sec:evol_and_setup}

Both the Einstein and the full set of filtered GRMHD evolution equations, including all the gradient SGS terms, can be found in~\cite{vigano20,aguilera20}. Following those works, we include only the SGS term (i.e., $\tau^{ki}_{M}$) appearing on the induction equation, that in a flat spacetime can be written as
\begin{eqnarray}
\partial_t {B}^i + \partial_k [ 
B^{i} v^{k} - B^{k} v^{i}  - {\tau}^{ki}_{M} ] = 0~,
\tau^{ki}_{M} = -\frac{\Delta x^2}{24} {\cal C_M} \, H_{M}^{ki}\nonumber
\label{evol_B_sgs}
\end{eqnarray}
where the explicit expressions for the tensor $H_{M}^{ki}$ in terms of field gradients can be found in \cite{aguilera20}. As we found in that study, the value  ${\cal C_M}=8$ reproduces the magnetic field amplification more accurately for our numerical schemes, and we will set the same value for  all our simulations in this work. Note that less dissipative numerical schemes, or higher resolution simulations, would require smaller values of ${\cal C_M}$ closer to the theoretical expectation ${\cal C_M} \simeq 1$.

As in our previous works, we use the code {\sc MHDuet}, generated by the platform {\sc Simflowny} \citep{arbona13,arbona18,palenzuela21b} and based on the {\sc SAMRAI} infrastructure \citep{hornung02,gunney16}. {\sc MHDuet} employs at least fourth-order-accurate operators to discretize both the spatial and time derivatives, with high-resolution shock-capturing method for the fluid based on the Lax-Friedrich flux splitting formula \citep{shu98} and the fifth-order reconstruction method MP5 \citep{suresh97}.
A complete assessment of the implemented numerical methods can be found in \cite{palenzuela18}. We employ the same hybrid EoS as in~\cite{palenzuela21} for the evolution, with a cold contribution given by a tabulated polytrope fit to the APR4 zero-temperature EoS~\citep{read09}, and thermal effects modeled by the ideal gas EoS with adiabatic index $\Gamma_{\rm th}= 1.8$.

The conversion from the conserved fields to the primitive ones is performed by using the robust procedure given in~\cite{kastaun20}. To minimize further failures in the very tenuous regions outside the star, we impose a minimum density of $6.2 \times 10^{5}~\rm{g~cm^{-3}}$, with the regions having such values referred hereafter as atmosphere. Moreover, we apply the SGS terms only in regions where the density is higher than $6.2 \times 10^{13}~\rm{g~cm^{-3}}$ in order to avoid spurious effects near the stellar surface. Since the remnant's maximum density is above $10^{15} ~\rm{g~cm^{-3}}$, the SGS model is applied only in the most dense regions of the star.

\begin{figure*}
	\centering
	\includegraphics[width=0.285\linewidth, trim={0 4.5cm 0 0}, clip]{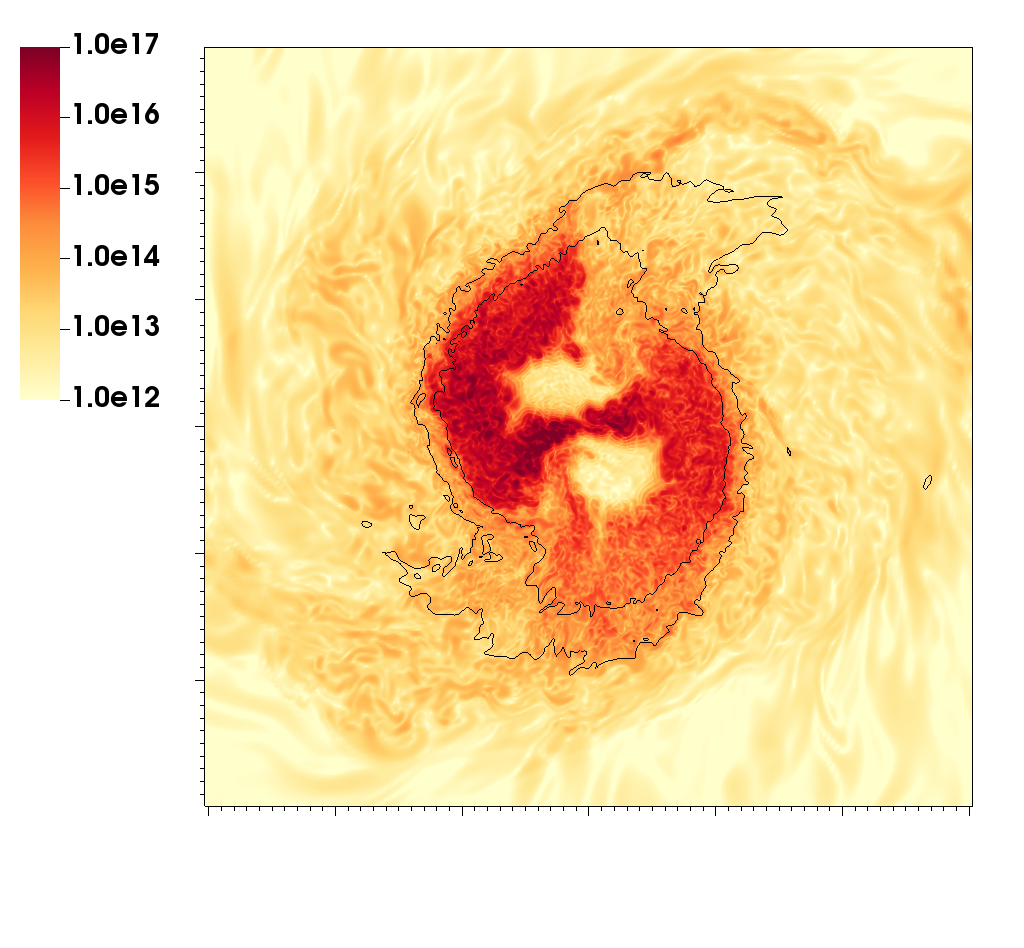}
	\includegraphics[width=0.233\linewidth, trim={6.5cm 4.5cm 0 0}, clip]{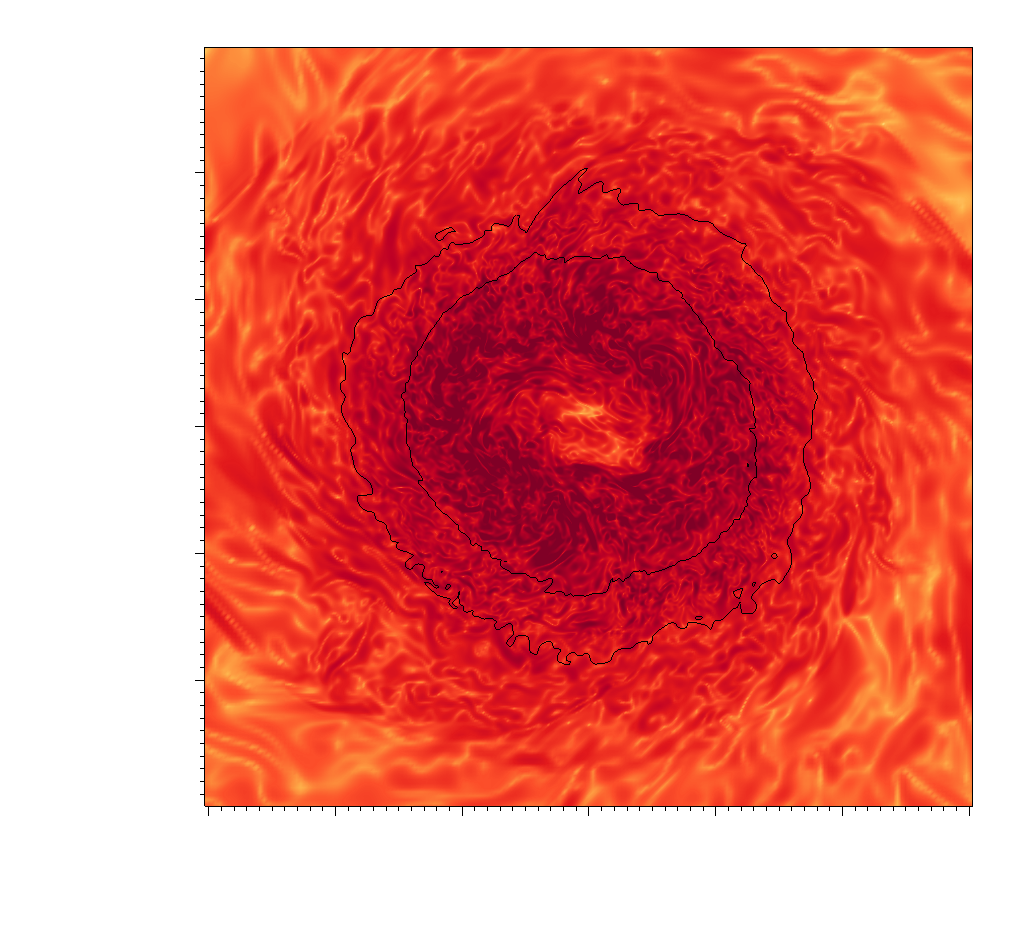}
	\includegraphics[width=0.233\linewidth, trim={6.5cm 4.5cm 0 0}, clip]{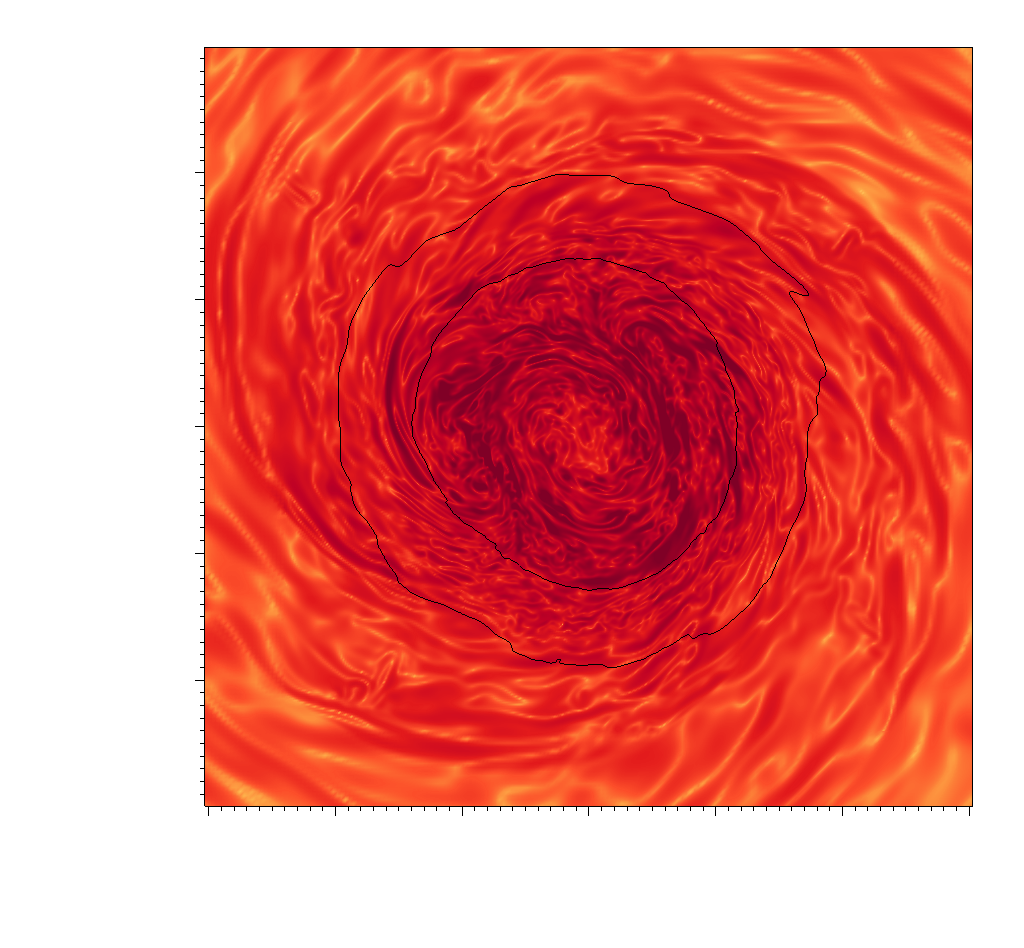}
	\includegraphics[width=0.233\linewidth, trim={6.5cm 4.5cm 0 0}, clip]{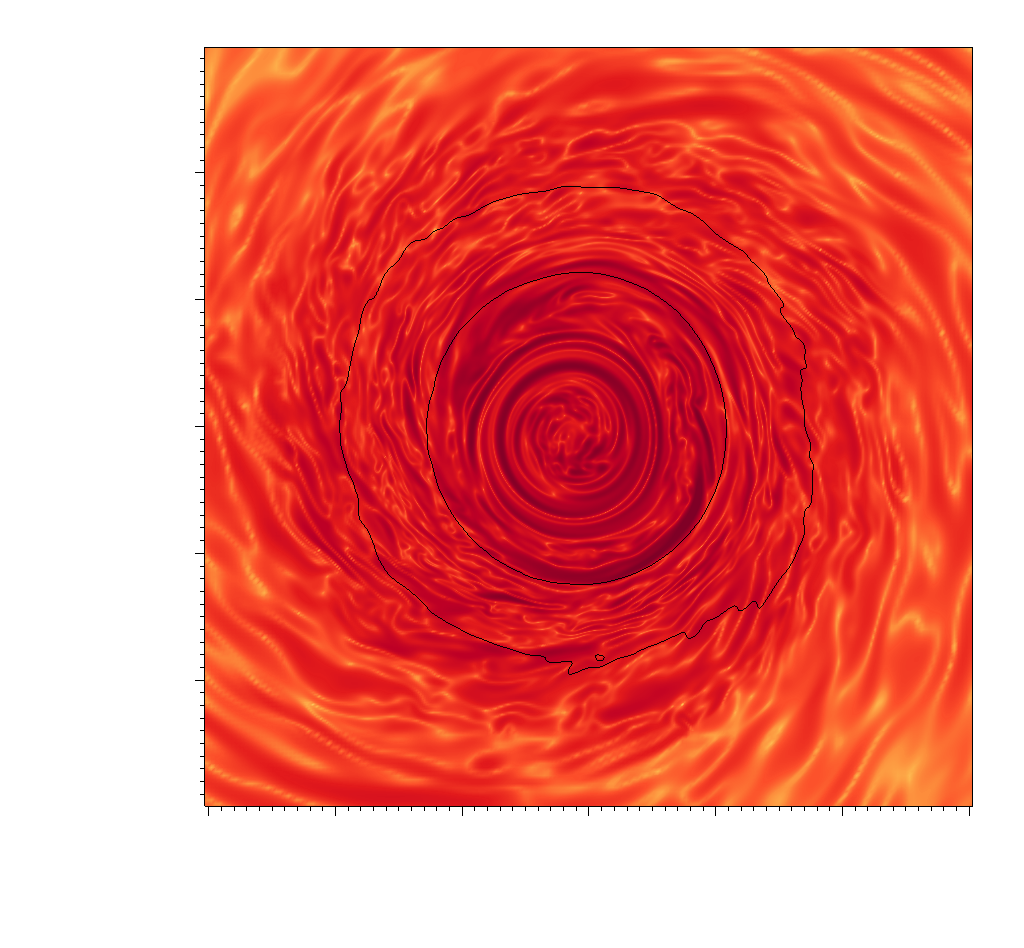}\\
	\includegraphics[width=0.285\linewidth, trim={0 4.5cm 0 0}, clip]{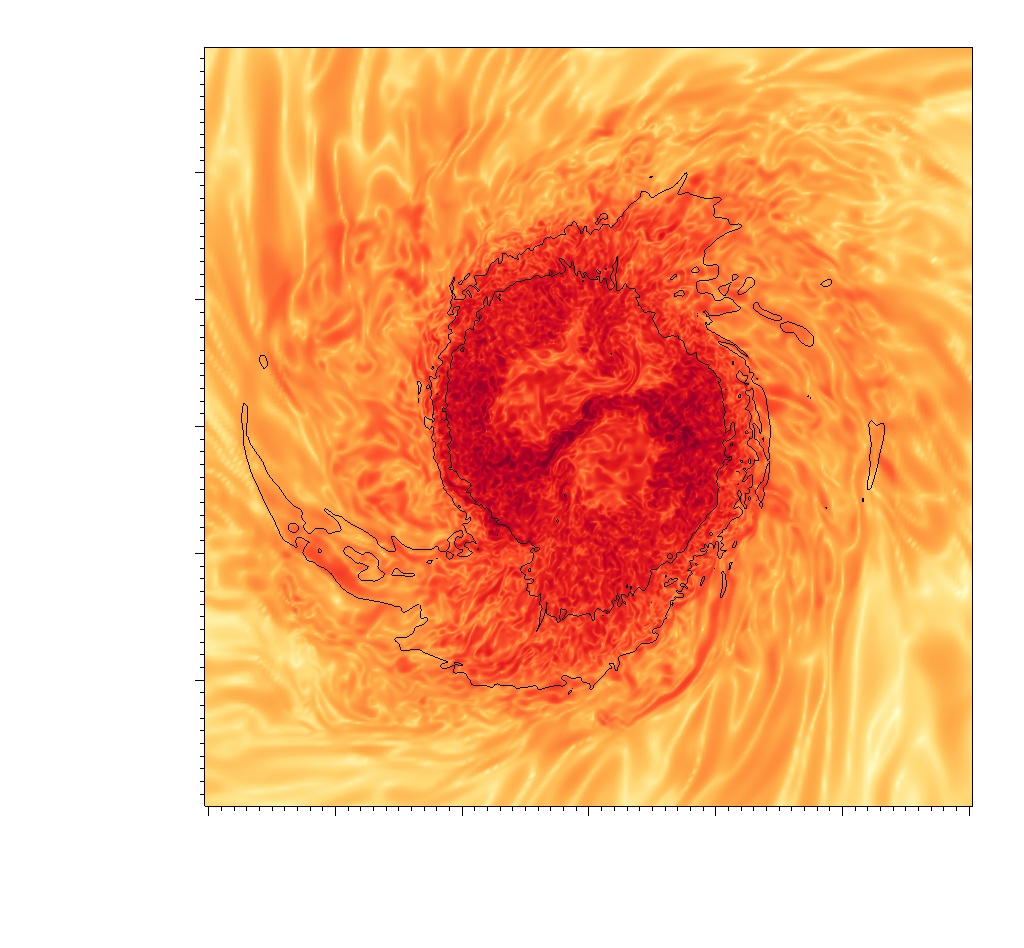}
	\includegraphics[width=0.233\linewidth, trim={6.5cm 4.5cm 0 0}, clip]{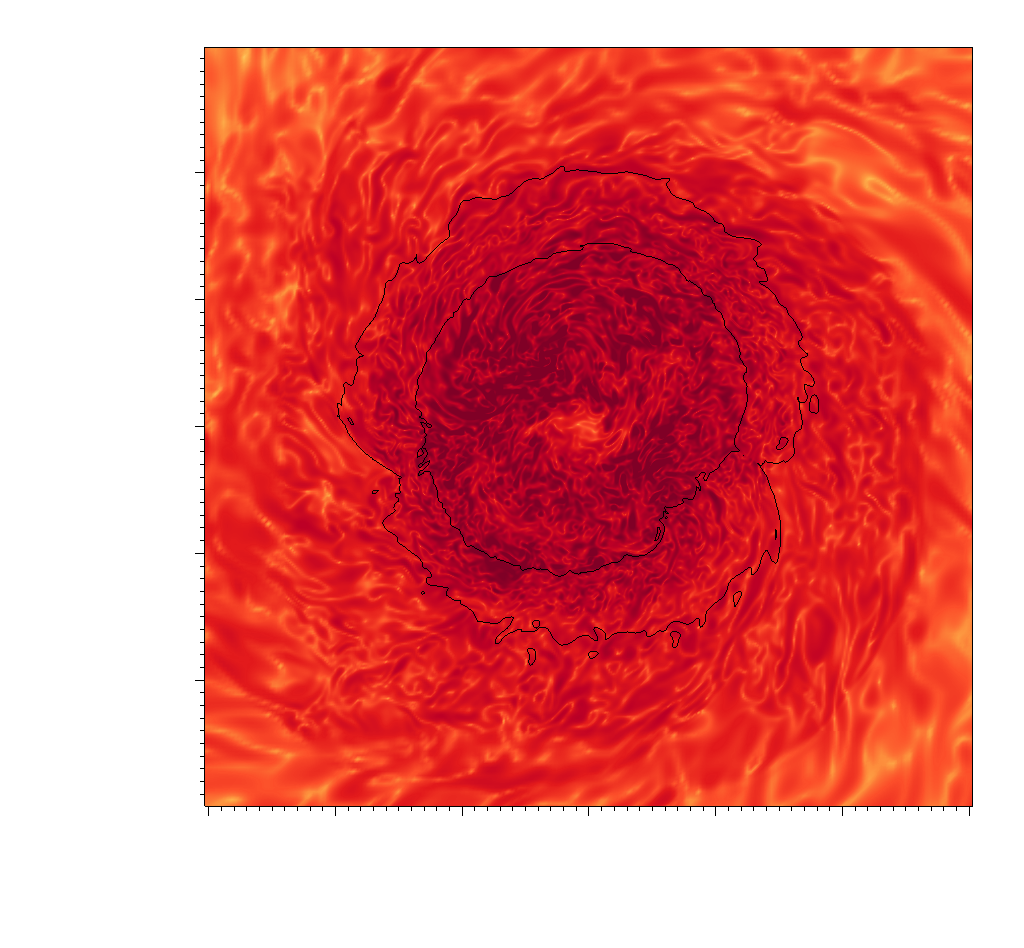}
	\includegraphics[width=0.233\linewidth, trim={6.5cm 4.5cm 0 0}, clip]{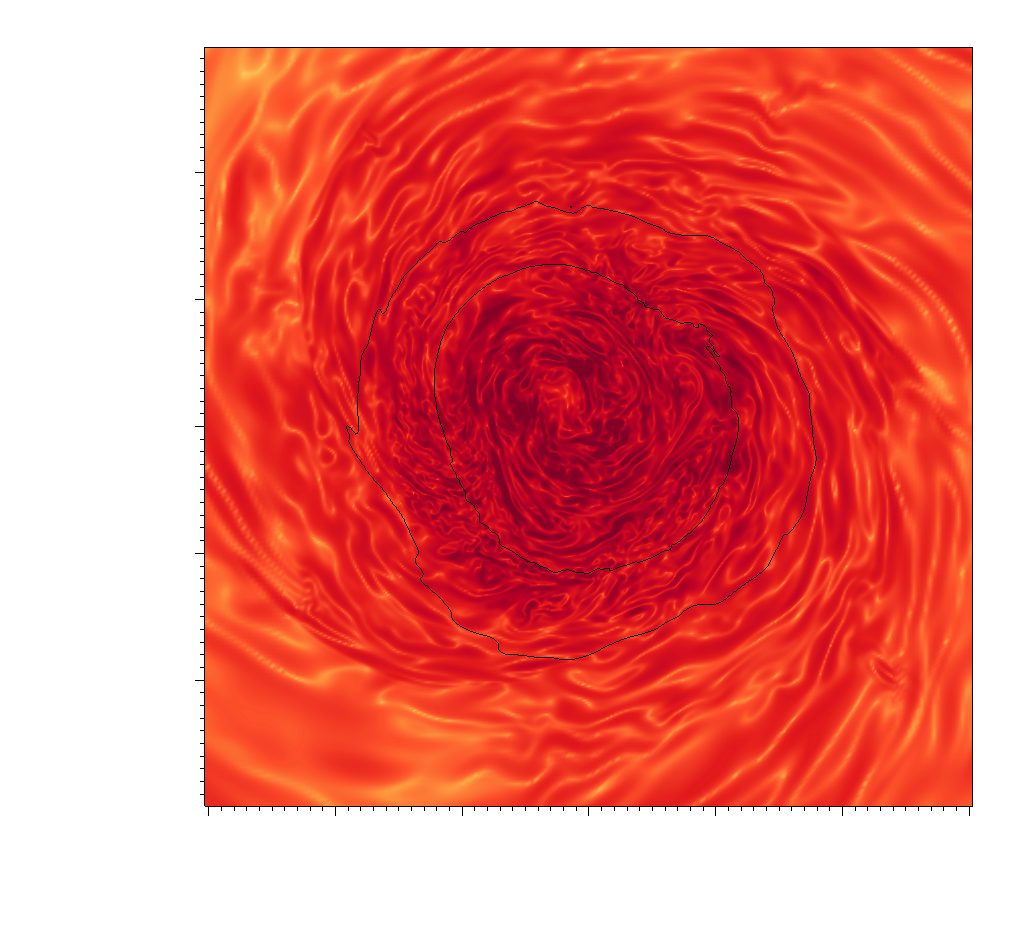}
	\includegraphics[width=0.233\linewidth, trim={6.5cm 4.5cm 0 0}, clip]{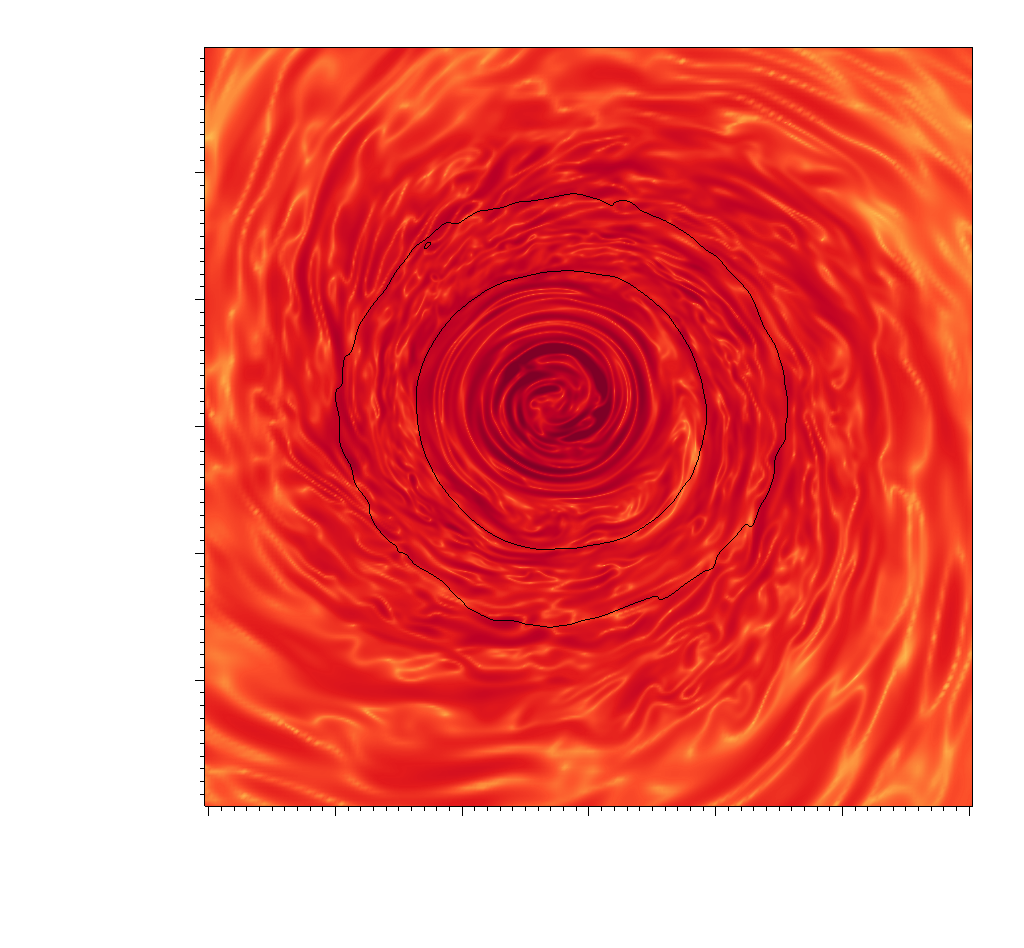}\\
	\includegraphics[width=0.285\linewidth, trim={0 4.5cm 0 0}, clip]{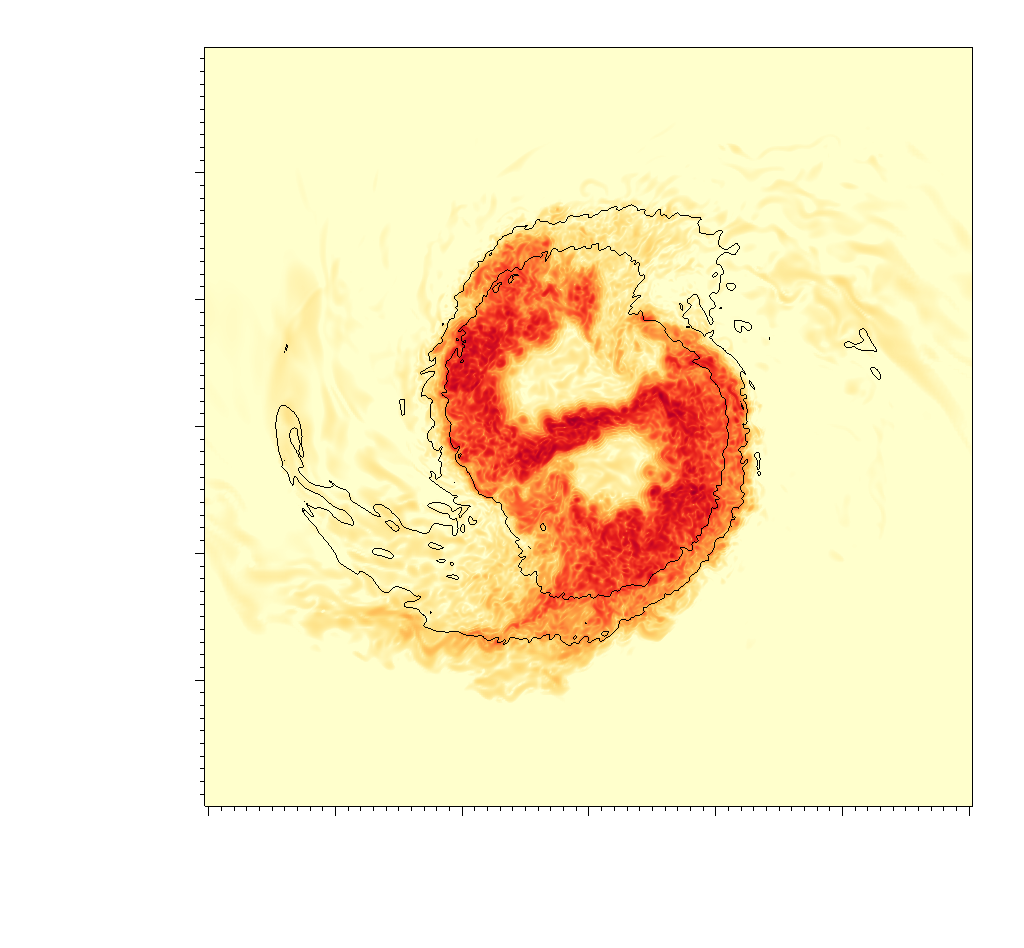}
	\includegraphics[width=0.233\linewidth, trim={6.5cm 4.5cm 0 0}, clip]{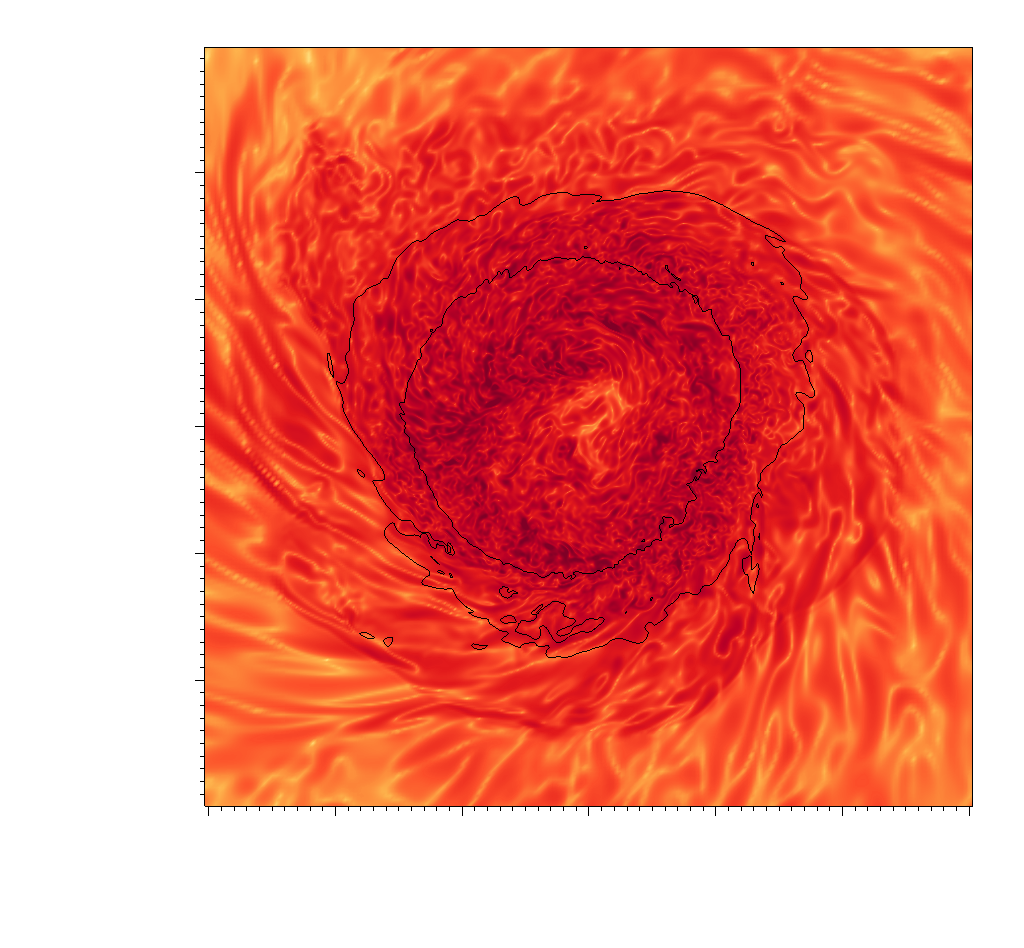}
	\includegraphics[width=0.233\linewidth, trim={6.5cm 4.5cm 0 0}, clip]{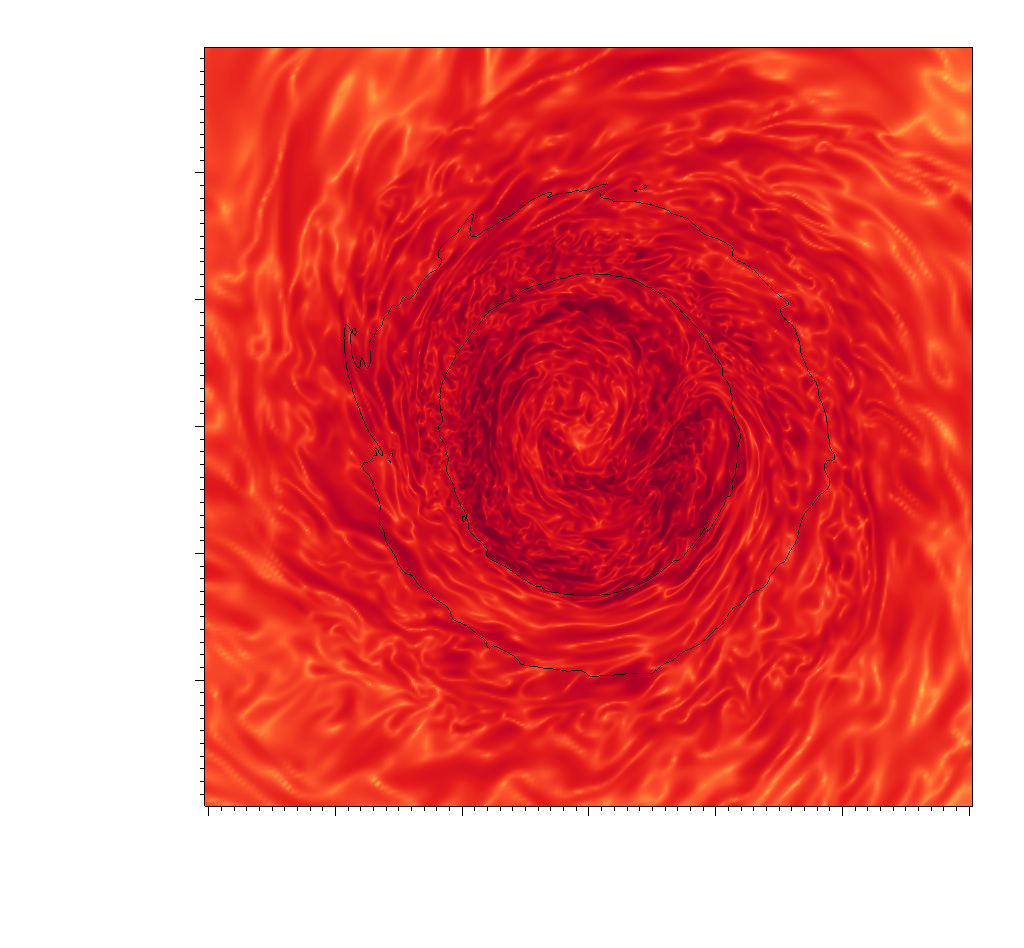}
	\includegraphics[width=0.233\linewidth, trim={6.5cm 4.5cm 0 0}, clip]{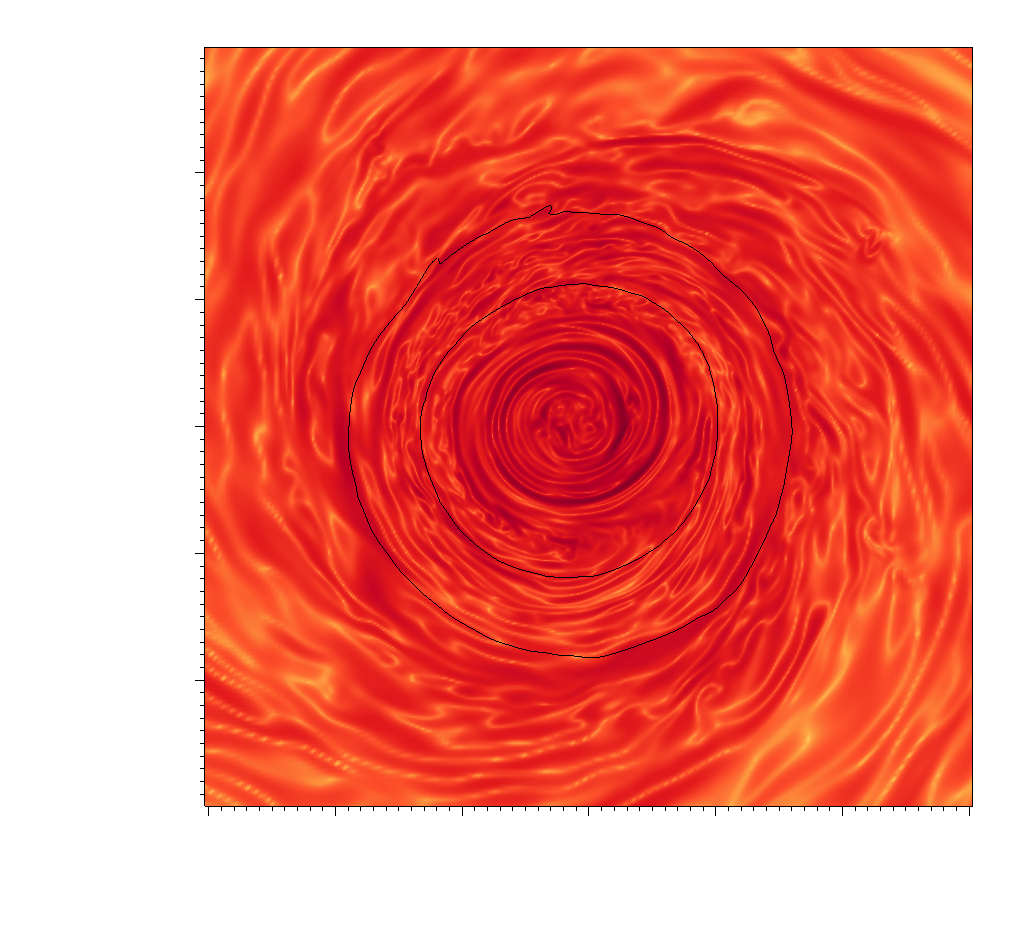}\\
	\includegraphics[width=0.285\linewidth, trim={0 2cm 0 0}, clip]{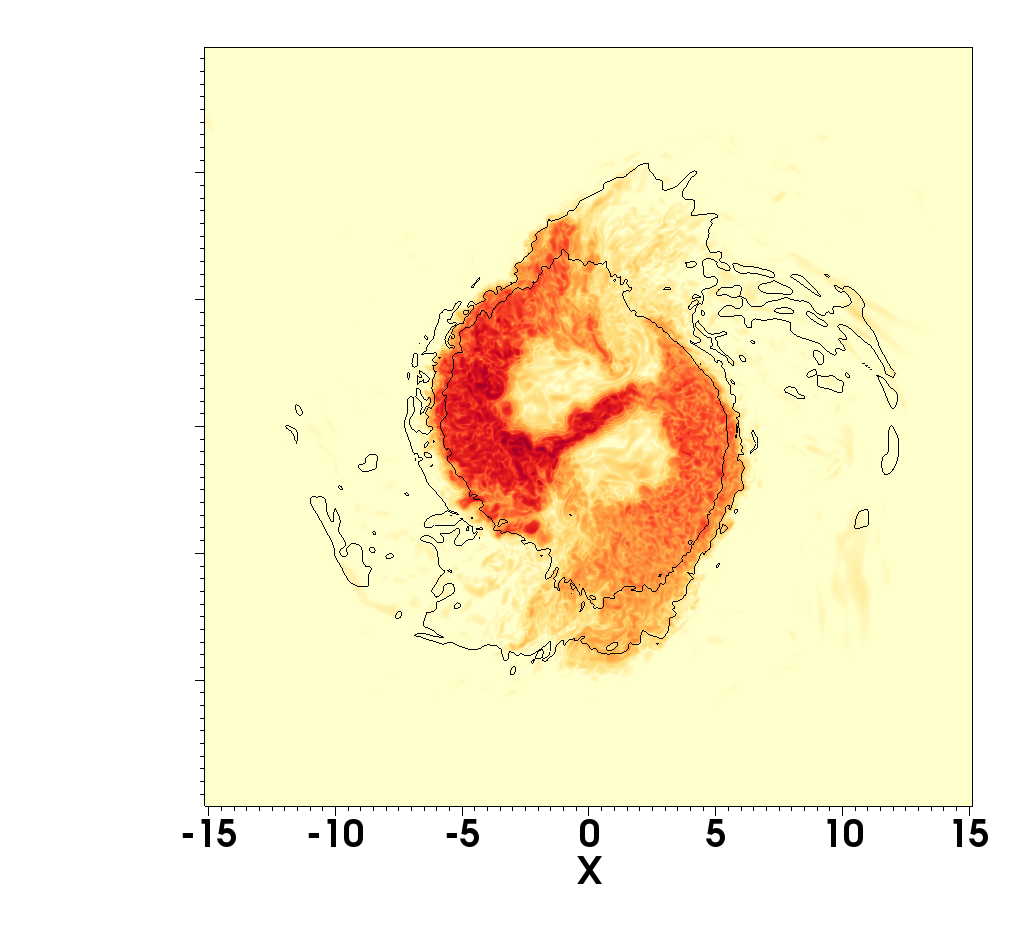}
	\includegraphics[width=0.233\linewidth, trim={6.5cm 2cm 0 0}, clip]{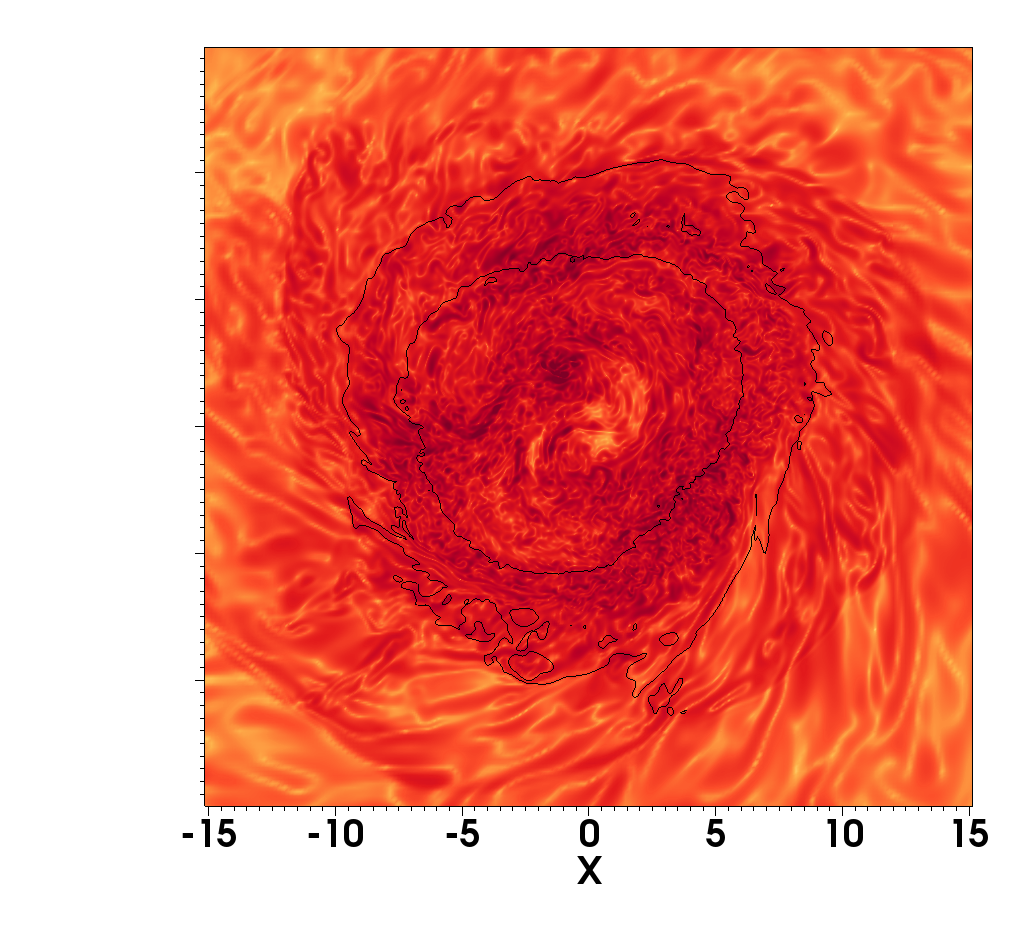}
	\includegraphics[width=0.233\linewidth, trim={6.5cm 2cm 0 0}, clip]{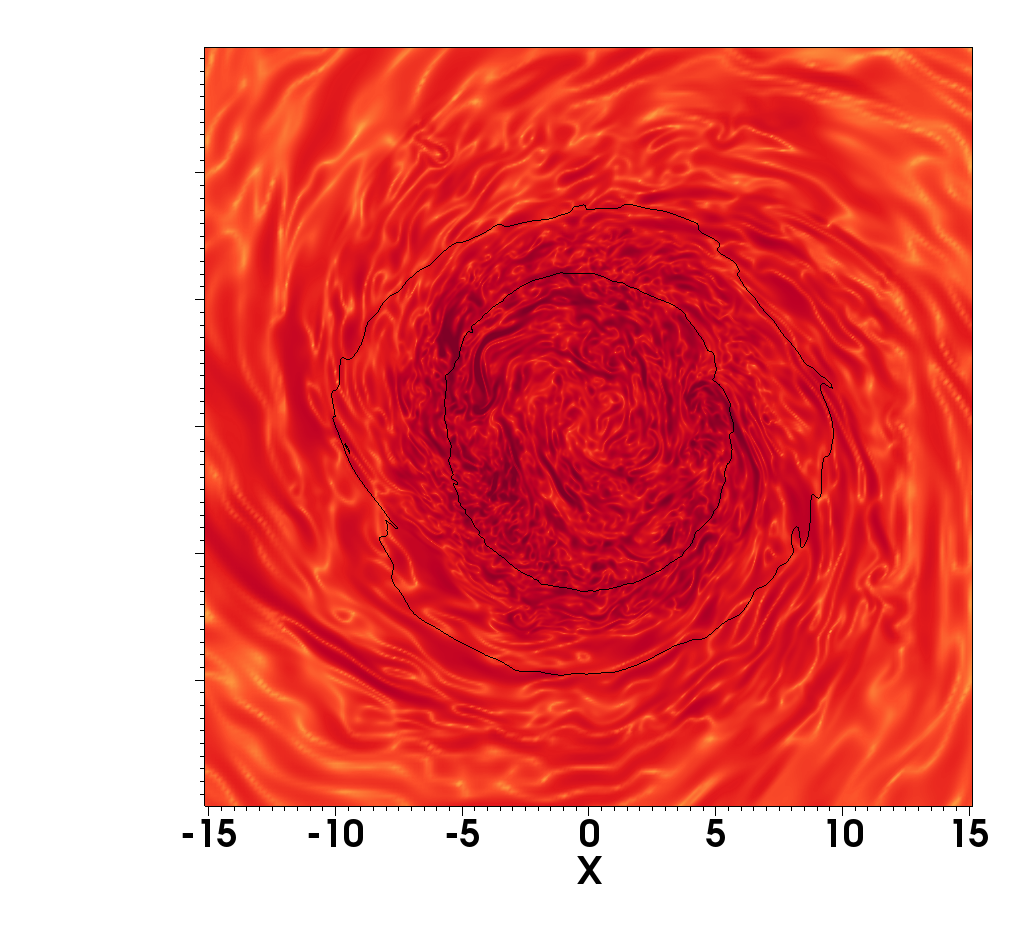}
	\includegraphics[width=0.233\linewidth, trim={6.5cm 2cm 0 0}, clip]{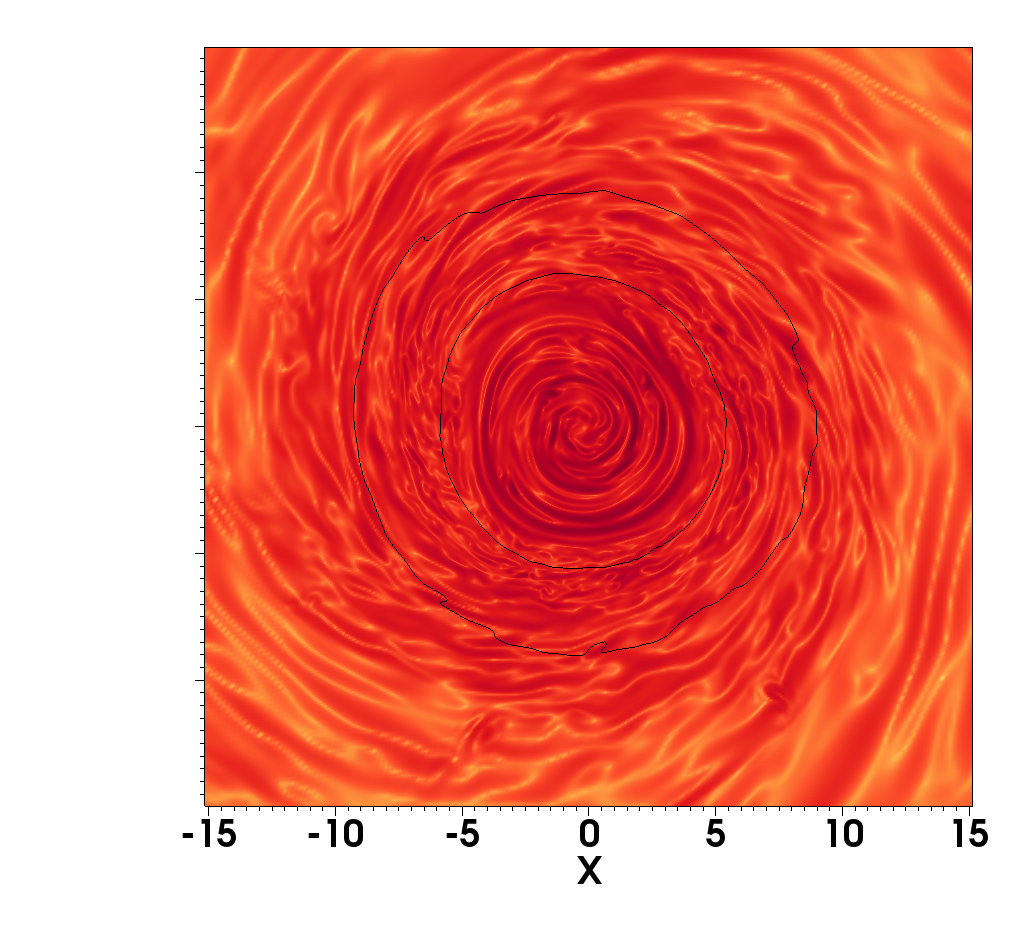}\\
	\caption{\textit{Magnetic field evolution}. Values of the magnetic intensity $|\vec{B}|$ in the orbital plane for: from top to bottom, {\ttfamily Dip}, {\ttfamily BHigh}, {\ttfamily Misal} and {\ttfamily Mult} simulations at, from left to right, $t=\{2,5,10,20\}$ ms after the merger. Outer and inner black lines mark the contours $\rho = 10^{13}$ and $10^{14}\ \rm{g/cm^{-3}}$, respectively. The length are given in geometrical units (corresponding to 1.47 km).}
	\label{fig:slices_B2}
\end{figure*}

The initial data is created with the {\sc Lorene} package~\cite{lorene}, using the same tabulated polytropic EoS described above. We consider an equal-mass BNS in quasi-circular orbit with an irrotational configuration. The total mass of the system is $M=2.7~M_{\odot}$ and the initial  separation is $45$km, corresponding to an initial angular frequency of $1775\ \rm{rad~s^{-1}}$. 

The binary is solved in a cubic domain of side
$\left[-1204,1204\right]$ km. The inspiral is fully covered by $7$ Fixed Mesh Refinement (FMR) levels, each being a cube doubling the resolution of the previous one, and one Adaptive Mesh Refinement (AMR) level, achieving a maximum resolution of $60$ m in a domain covering at least the bulk of the remnant.

For each star, we consider a commonly-used initial axially-symmetric magnetic field, confined in the region where the fluid pressure $P$ is larger than a value $P_{cut}$, set to 100 times the atmospheric pressure. The azimuthal ($\phi$) component of the vector potential has a radial dependence $A_ {\phi} \propto r^2 (P - P_{cut})$, where $r$ is the distance from the center of the star. We have considered four initial configurations that differ among themselves in the intensity and the co-latitude ($\theta$) dependence of the magnetic field, as follows (see also Table~\ref{tab:models}):
\begin{itemize}
	\item Aligned Dipole-like ({\ttfamily Dip}): 
	A very ordered (large scale) poloidal field $A_\phi = A_0 r^2 \sin^2\theta  (P - P_{cut})$, similar to a dipole (which would go $\propto \sin\theta$), with the magnetic moment aligned to the orbital axis, and a normalization value $A_0$ such that the maximum intensity (at the centre) is $10^{12}~\rm{G}$, orders of magnitude lower than the large initial fields of other simulations (e.g., \cite{kiuchi15,ruiz16,kiuchi18,ciolfi2019,ciolfi2020collimated,ruiz2020}). 
	\item Highly magnetized ({\ttfamily BHigh}): The same as the above {\ttfamily Dip} model, except that the intensity is 1000 times larger, reaching a maximum of $10^{15}$ \rm{G}.
	\item Misaligned dipole ({\ttfamily Misal}): The same as the {\ttfamily Dip} model, except that the magnetic moment is orthogonal to the orbital axis.
	\item Multipole ({\ttfamily Mult}): A more complex topology containing high multi-polar structures, with $A_ {\phi} \propto r^2 \sin^6\theta\left(1+\cos\theta\right) (P - P_{cut})$, with the same maximum intensity as {\ttfamily Dip}.
\end{itemize}

\begin{table}[ht]
\begin{tabular}{ |c|c|c|c| }			
	\hline
	Case
	& Max $B$
	& Orbit-magnetic
	& Meridional\\
	& (G) & misalignment (degrees) & topology
	\\ \hline
	{\tt Dip} & $10^{12}$ & 0 & Dipole-like \\
	{\tt BHigh} & \cellcolor{gray!25}$10^{15}$ & 0 & Dipole-like \\
	{\tt Misal} & $10^{12}$ & \cellcolor{gray!25}90 & Dipole-like \\
	{\tt Mult} & $10^{12}$ & 0 & \cellcolor{gray!25}Multipole \\
	\hline
\end{tabular}
\caption{{\em Configuration of the simulations.} We indicate the initial values of the maximum intensity of the magnetic field, the angular misalignment between the orbital and magnetic field axes, and the initial topology of the magnetic field. Cells remarked in grey correspond to the differences with respect to the reference model {\ttfamily Dip}.}
\label{tab:models}
\end{table}

\section{Results}\label{sec:results}

\begin{figure}
	\centering
	\includegraphics[width=0.7\linewidth]{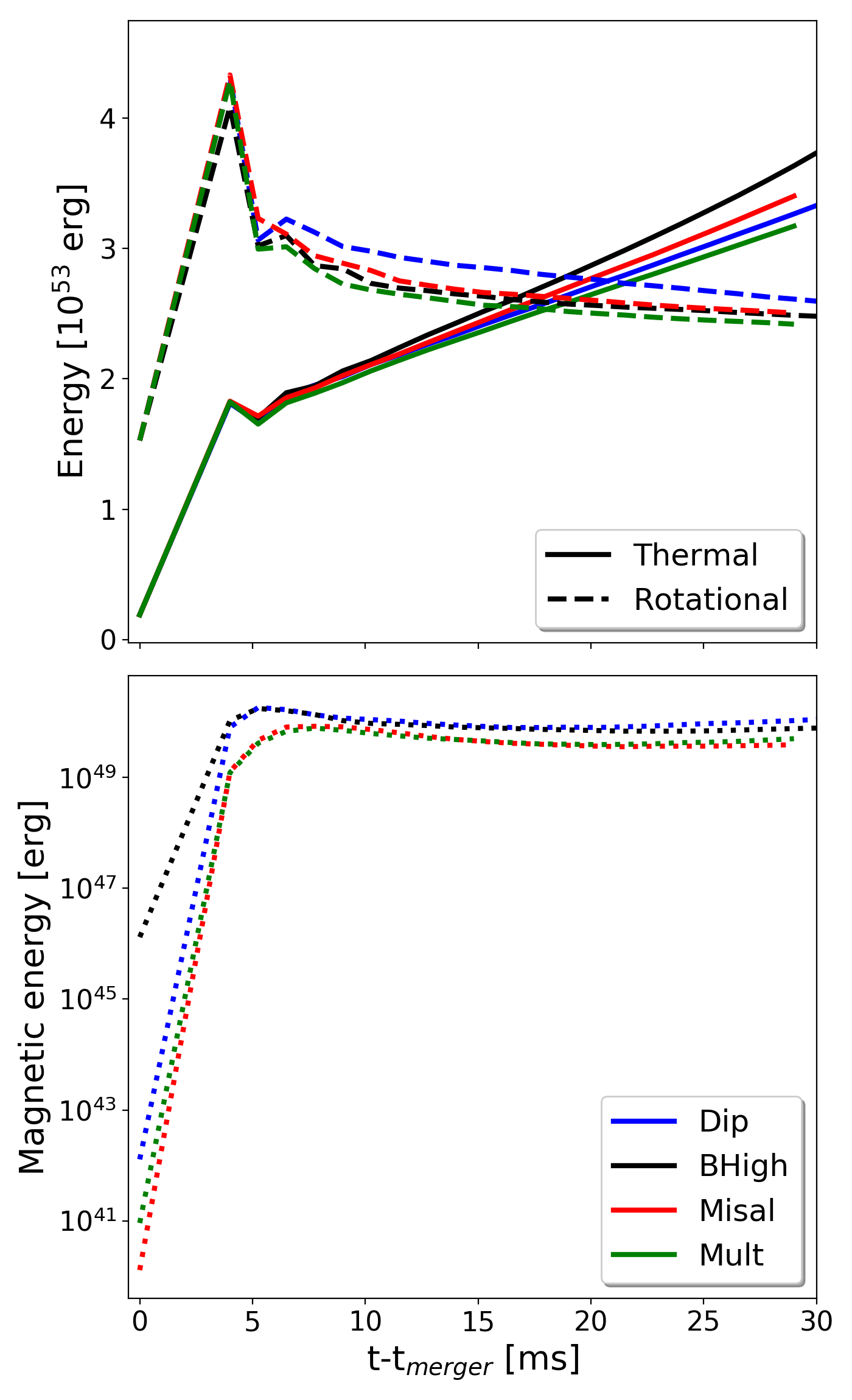}
	\caption{\textit{Energy evolution}. (Top) Rotational (dashed lines) and thermal (solid lines) energies, integrated over the whole dominion, for different simulations as a function of time. (Bottom) Magnetic energy for the same simulations.}
	\label{fig:integrals}
\end{figure}

\begin{figure}
	\centering
	\includegraphics[width=0.7\linewidth]{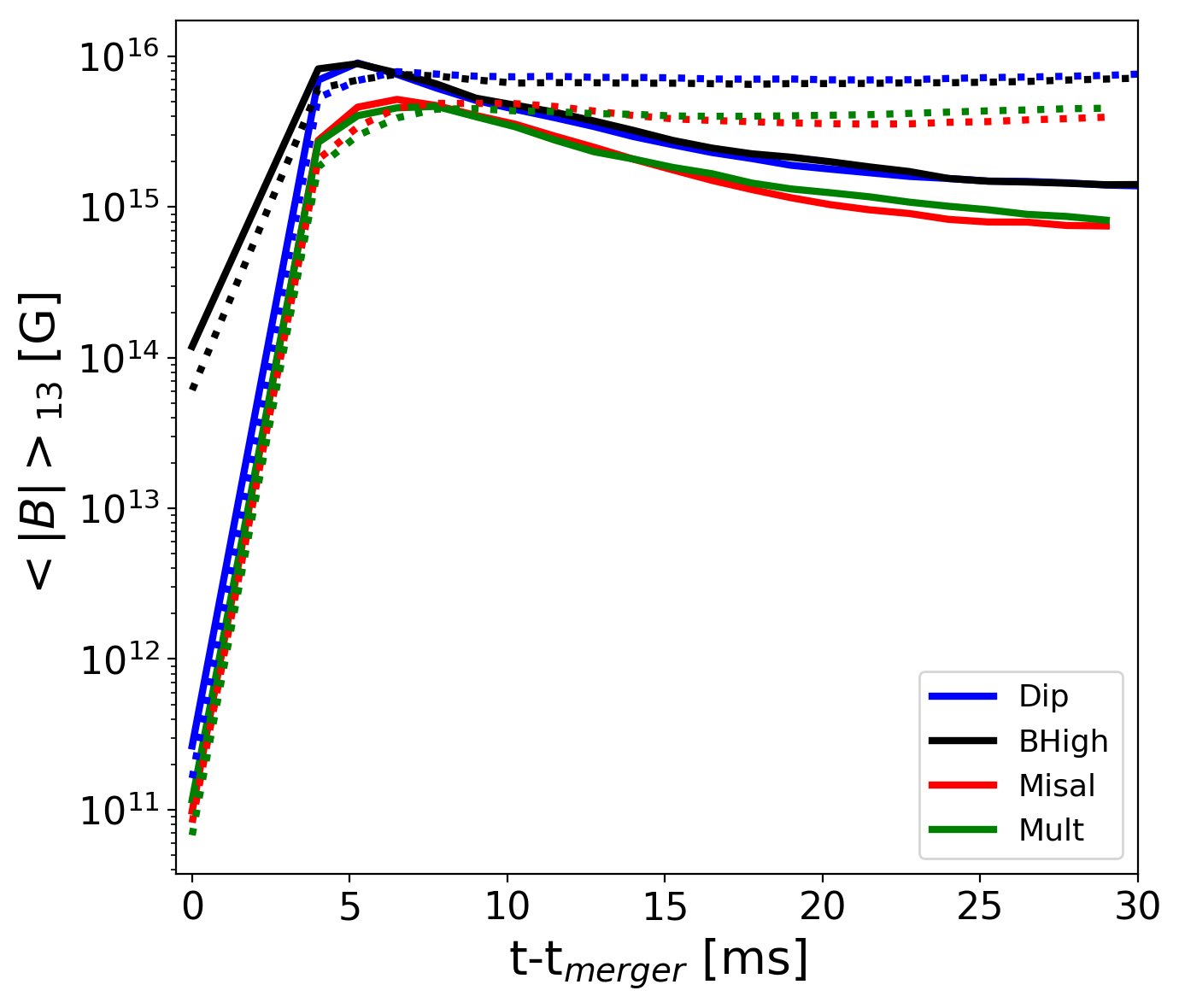}
	\caption{\textit{Average intensity of magnetic field components}. Average intensity evolution of the poloidal (solid lines) and toroidal (dashed lines) components of the magnetic field for all cases in the bulk of the remnant where $\rho\geq10^{13}\ \rm{g~cm^{-3}}$.}
	\label{fig:averagebbpolbtor}
\end{figure}

Our initial binary system evolves for $5$ orbits before merging. The merger produces a differentially rotating remnant that relaxes to an hypermassive-neutron star (HMNS) in a few milliseconds. Before the merger occurs (hereafter, $t=0$), we set a magnetic field topology on each star corresponding to one of the cases described before. Therefore, we have then considered four different simulations, as summarized in Table~\ref{tab:models}.

Fig.~\ref{fig:slices_B2} displays some snapshots, in the orbital plane $z=0$, of the {\ttfamily Dip}, {\ttfamily BHigh}, {\ttfamily Misal} and {\ttfamily Mult} simulations (from top to bottom) at $t=\{2,5,10,20\}$ ms (from left to right) after the merger. The orange scale represents the intensity of the magnetic field, while the two black thin lines are mass density contours corresponding to $\rho = 10^{14}~\rm{g~cm^{-3}}$ (inner line) and $10^{13}~\rm{g~cm^{-3}}$ (outer line). The shape of the remnant varies at the initial times, but clearly restructures itself at later ones, approaching an almost axisymmetric structure. There we can also see the KHI appearing at the merging layers, amplifying local values of MF up to a maximum of $10^{17}$ G. Thus, MF changes from fully turbulent to partially ordered at $\sim 20$ ms after the merger, where we can see the systematic formation of azimuthal/spiraling filaments. This is due to the winding that from now on rule the rising of the magnetic field (see \cite{palenzuela21} for an in-depth discussion about the mechanisms contributing to the amplification).

In Fig.~\ref{fig:integrals} we represent the evolution of the volume-integrated thermal energy (top, solid line), kinetic rotational (top, dashed line) and magnetic energy (bottom) for the four models. We can see the energies of all cases rise soon after merger, at the expense of the large gravitational energy available. For all cases, the thermal energy still rises monotonically after the merger while the rotational one is losing energy, becoming all remnants objects with higher temperature but rotating more slowly. We obtain comparable values between all cases at $30$ ms after the merger for the rotational energy. For the thermal one, differences are around a factor of $\sim1.2$ between {\ttfamily BHigh} and {\ttfamily Mult} at the same time.

After about 5 milliseconds, the magnetic energy growth saturates, with a maximum factor difference of $\sim3$ between {\ttfamily Dip} and {\ttfamily Misal}. The KHI that appears during the merger between the two stars, possibly combined with the Rayleigh-Taylor instability near the surface, is the responsible of the amplification of the magnetic energy, which for all cases (except by the {\ttfamily BHigh}) increases $10$ orders of magnitude, from $10^{40}\ \rm{erg}$ to $10^{50}\ \rm{erg}$. For the {\ttfamily BHigh} case, the initially large value implies a smaller amplification by $4$ orders of magnitude from $10^{46}\ \rm{erg}$ to a similar value of $10^{50}\ \rm{erg}$. Overall, differences in energies lie within a factor $\sim3$, much less than both the orders-of-magnitude differences in the initial magnetic energy and across the evolution.

\begin{figure*}
	\centering
	\includegraphics[width=\linewidth]{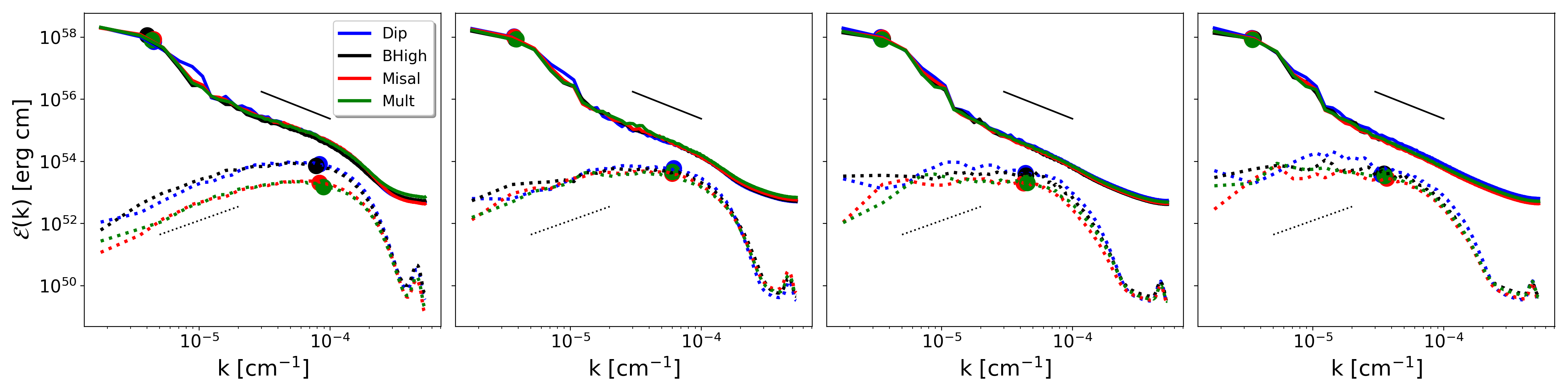}
	\caption{ \textit{Energy spectra}. Top: Kinetic (solid) and magnetic (dotted) energy spectra for different configurations as a function of the wavenumber at, from left to right, $t=\{5,10,20,30\}$ ms after the merger. The solid thin black line corresponds to the Kolmogorov slope, while the dotted one belongs to the Kazantsev slope. The dots are the wavenumbers which contains, in average, most the energy of each spectra.}
	\label{fig:spectra_all}
\end{figure*}

\begin{figure*}
	\centering
	\includegraphics[width=\linewidth]{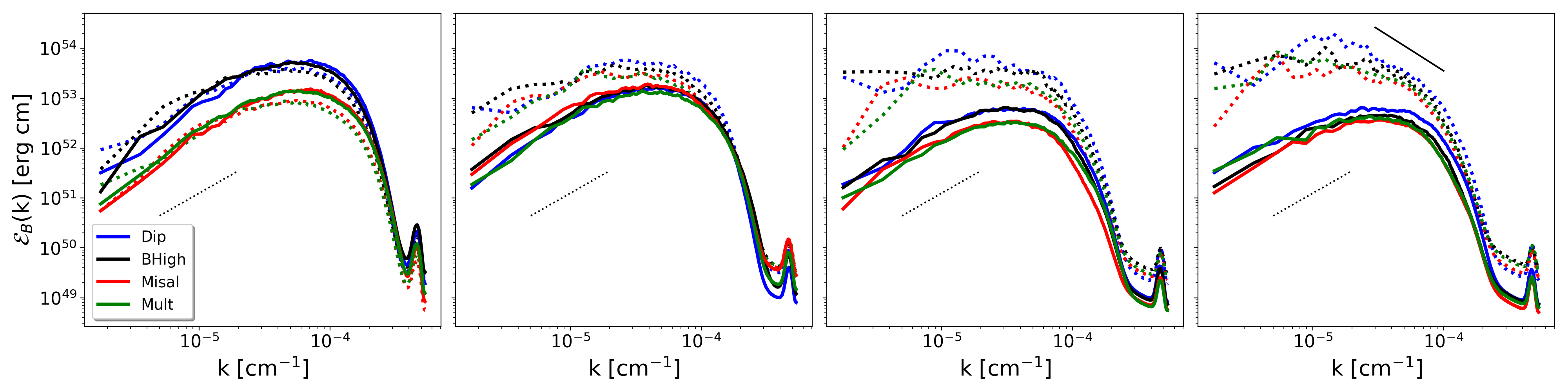}
	\caption{\textit{Magnetic energy spectra}. Poloidal (solid) and toroidal (dotted) components of the magnetic spectra for different configurations as a function of the wavenumber at, from left to right, $t=\{5,10,20,30\}$ ms after the merger.}
	\label{fig:tor_pol_spectra_all}
\end{figure*}

Fig.~\ref{fig:averagebbpolbtor} shows the  intensity of the poloidal (solid lines) and toroidal (dashed lines) components of the magnetic field, as a function of time, averaged in the bulk of the remnant. There we can see the amplification for both components due to the KHI in the firsts $5$ milliseconds after the merger. What is important to remark here is that: (i) during the exponential growth, both components are similar due to the isotropic character of turbulence, and (ii) after the rising, the toroidal component is the dominant one, as its shape is practically the same as the magnetic energy plot from Fig.~\ref{fig:integrals}. The toroidal component, in the saturation phase, maintains its value merely constant for all cases, close to $10^{16}~\rm{G}$. The poloidal component, on the other hand, is slowly decreasing, probably due to energy transfer to the toroidal component, getting values around $10^{15}~\rm{G}$. For the toroidal component of the magnetic field we can see almost the same behaviour for all cases where, again, the misaligned and the multipolar ones are below the others by a factor $\sim3$. Differences are less notorious when focusing on the poloidal component of the magnetic field, where at the beginning the same cases (i.e. misaligned and multipolar) do not rise as much as the others but at $30$ ms after the merger the differences are only about a factor $\sim2$. 

Besides the volume-integrated quantities, we can analyze the evolving spectral energy distribution ${\cal E}(k)$ (for details on how it is calculated, see appendix of \cite{vigano20}). This will allow us to see whether the different scales of the problem behave similar or not between the cases we are considering. The spectral distribution of the kinetic (solid lines) and magnetic (dotted lines) energies, as a function of the wavenumber $k$, is displayed in Fig.~\ref{fig:spectra_all}. The four plots correspond to, from left to right, $t=\{5,10,20,30\}$ ms after the merger. As a reference, Kolmogorov ($k^{-5/3}$, thin solid line) and Kazantsev ($k^{3/2}$, thin dotted line) slopes are also included in all plots. Large dots indicate the spectra-weighted average of the wavenumber, ${\bar k}=\frac{\int_k [k {\cal E}(k)]}{\int_k {\cal E}(k)}$, which gives the typical size ${\bar \lambda}= 2 \pi/{\bar k}$ of either the fluid or the magnetic (${\bar \lambda}_B$) structures.

For all times represented, the kinetic energy spectra behave in the same way for all cases (having a Kolmogorov slope in the inertial range), regardless of the scale we are considering. For the magnetic energy spectra, they initially follow the Kazantsev slope up to the numerical dissipative scale (intrinsically set by the discretization scheme). This is a property of the kinematic phase of the dynamo, until the dynamo approaches saturation at small scales (large $k$). At $t=10\ \rm{ms}$ all cases have roughly the same amount of magnetic energy spectra. At $t=20\ \rm{ms}$ after the merger, small differences begin to appear, although they may be in part due to stochastic variations. At $t=30\ \rm{ms}$ after the merger, such differences are less evident. Moreover, the amplification has saturated and magnetic spectra appear to be compatible with a Kolmogorov slope at intermediate scales.

We found that ${\bar \lambda}_B \sim 800$~m soon after the merger, increasing to almost $2$~km at $t=30$~ms. This confirm that larger coherent magnetic field structures are being formed in the remnant. Clearly, there is no significant difference (less than $7\%$ in ${\bar \lambda}_B$) between the simulations with different initial topologies considered here.

In Fig.~\ref{fig:tor_pol_spectra_all} we further analyze the magnetic energy spectra, identifying the contributions coming from the toroidal (dashed lines) and poloidal (solid) components. At $5$ ms after the merger, both components are similar for all simulations. As time passes, the poloidal component of all cases decreases two orders of magnitude while the toroidal one increases by about one order of magnitude. However, for all times, both components are still comparable among the different simulations. Notice that, comparing both components in different models, the differences are up to a factor $3$ only, much less than the relative differences and the overall changes in time. Finally, at $t=30$ ms, the slope of the toroidal component (the dominant contribution to the magnetic energy) approaches a Kolmogorov slope at intermediate scales. An interesting difference at these times is that, starting with a (unrealistically) high magnetic field, there is an excess of large-scale magnetic fields (low $k$), an effect probably due to the winding acting on the already quite organized field.

Finally, notice that the resemblance of integrated magnetic energy and the spectra for the different models point to a universality of the magnetic field not only in the bulk, but in the entire domain.

\section{Conclusions}\label{sec:conclusions}

We have studied the influence of different initial magnetic configuration on the evolution of BNS mergers, using high-resolution large-eddy simulations employed also in the accompanying paper~\cite{palenzuela21}, for which we refer for further details on the methods and amplification mechanisms at work. In particular, we have considered initial magnetic fields confined within the stars, varying the intensity, the magnetic moment misalignment with the orbital axis and the poloidal topology. Looking at the evolution of integrated energy and spectral distribution, we proved that the differences lie within a factor $3$ at most, which could be even smaller in more accurate simulations. This, then, ensures that the initial topology of the stars is not relevant at all because the small-scale turbulence induced in the remnant will erase any memory of realistic magnetic fields of $B \leq 10^{12}$G in only few milliseconds after the merger.
	
In this work we have explored only some of the infinite possible magnetic configurations. More choices could be explored, in particular: presence of a toroidal field; non-axisymmteric topology; magnetic field extended outside the stars; more complex, small-scale dominated configuration... However, the results shown here already suggest that the expected dependence on the initial topology is basically negligible, compared to other much more uncertain issues. Among the latter, we mention the importance of the numerical capability to resolve the small-scale magnetic amplification, the physics involved in the post-merger phase (neutrino transport, temperature-dependent equation of state...).

This universality of the magnetic field outcome after the merger sets serious doubts on how could we infer the initial magnetic field of the stars in a BNS merger through multimessenger astronomy. The only foreseeable possibilities are through the presence of a precursor electromagnetic signal before the merger,
or other kind of outflows that may appear during the first ms after the merger, which would have information mainly on the initial topology and intensity of the magnetic field~\citep{palenzuela13a,palenzuela13b,ponce14}.

The final message is that the commonly used simplification on the topology of the magnetic field is acceptable in BNS mergers, as far as the magnetic field is not too large and enough turbulence is developed to erase the seed and produce the correct spectra distribution. In those cases, the system will tend to be practically the same remnant regardless of its initial configuration. However, if one wants to focus on the realistic generation of a large-scale field in the post-merger, the rule-of-thumb is basically the following: be sure that the large-scale initial magnetic field is much smaller than the amplified average field that the numerical scheme is able to reproduce. 

\subsection*{Acknowledgements}

We thank Riccardo Ciolfi, Wolfgang Kastaun and Jay Vijay Kalinani for providing us the EoS and for the useful discussions. This work was supported by European Union FEDER funds, the Ministry of Science, Innovation and Universities and the Spanish Agencia Estatal de Investigación grant PID2019-110301GB-I00. DV is funded by the European Research Council (ERC) Starting Grant IMAGINE (grant agreement No. [948582]) under the European Union’s Horizon 2020 research and innovation programme. DV's work was also partially supported by the program Unidad de Excelencia María de Maeztu CEX2020-001058-M. The authors thankfully acknowledge the computer resources at MareNostrum and the technical support provided by Barcelona Supercomputing Center (BSC) through Grant project Long-LESBNS by the $22^{nd}$ PRACE regular call (Proposal 2019215177, P.I. CP and DV).

\bibliographystyle{utphys}
\bibliography{turbulence}

\end{document}